\documentclass[12pt,a4paper]{article}

\usepackage{amsmath,amssymb,amsfonts,bm,graphicx,xspace}

\usepackage{enumerate}
\usepackage{cite}
\usepackage{xspace}
\usepackage{ifthen}
\usepackage{url}
\usepackage{booktabs}
\usepackage{multirow}
\usepackage{subcaption}
\usepackage{marginnote}
\usepackage{lineno}
\usepackage{amsmath,bm,graphicx}
\usepackage[utf8]{inputenc}
\usepackage{xcolor}

\newcommand{\pythia}{P\protect\scalebox{0.8}{YTHIA}\xspace}

\renewcommand{\eqref}[1]{eq.~(\ref{#1})\xspace}

\newcommand{\fig}[1]{\ref{#1}}
\newcommand{\figref}[1]{figure~\fig{#1}}

\newcommand{\citeref}[1]{ref.~\cite{#1}}

\newcommand{\sect}[1]{\ref{#1}}
\newcommand{\sectref}[1]{section~\sect{#1}}

\def\text{\mathrm}
\def\eg{\emph{e.g.}\xspace}
\def\ie{\emph{i.e.}\xspace}

\def\mrm#1{\mathrm{#1}}

\def\epem{$\mrm{e}^+\mrm{e}^-$\xspace}
\def\ep{$\mrm{ep}$\xspace}

\def\pp{\ensuremath{\mrm{pp}}\xspace}

\def\ep{\ensuremath{\mrm{ep}}}

\def\kappaS{\ensuremath{\kappa}}

\newcommand{\diff}[1]{\ensuremath{d#1}}
\newcommand{\dtwo}[1]{\ensuremath{d^2#1}}

\def\ColText(#1,#2)[#3]#4{\Text(#1,#2)[#3]{#4}}

\def\showcommentsflag{0}
\newcommand{\showcomments}{\def\showcommentsflag{1}}

\newcounter{commentcounter}%
\definecolor{armygreen}{rgb}{0.29, 0.33, 0.13}
\newcommand{\Red}[1]{\textcolor{red}{#1}}
\newcommand{\Black}[1]{\textcolor{black}{#1}}

\newcommand{\comment}[1]{\ifnum\showcommentsflag > 0%
        \addtocounter{commentcounter}{1}%
        {{\Red{\ensuremath{\ddagger^{\arabic{commentcounter}}}}\Black{}}}%
        \marginpar{\raggedright\tiny\it{{\Red{\ensuremath{\ddagger^{\arabic{commentcounter}}}}} {#1}\Black{}}}
        \fi%
}
\newcommand{\commentdel}[2]{\ifnum\showcommentsflag > 0%
        \Red{\sout{#1}}\comment{#2}%
        \fi
}
\newcommand{\commentadd}[2]{\ifnum\showcommentsflag > 0%
        \comment{#2}\Red{#1}%
        \else
        #1
        \fi
}
\newcommand{\commentchange}[3]{\ifnum\showcommentsflag > 0%
        \Red{\sout{#2}}\comment{#3}\Red{#1}%
        \else
        #1
        \fi
}
\newcommand{\nocomment}[1]{\ifnum\showcommentsflag > 0%
        {\tiny\it\Red{\{#1}\}}
        \fi%
}
\newcommand{\nocommentdel}[1]{\ifnum\showcommentsflag > 0%
        \Red{\sout{#1}}%
        \fi
}
\newcommand{\nocommentadd}[1]{\ifnum\showcommentsflag > 0%
        \Red{#1}%
        \else
        #1
        \fi
}

\newcommand{\nocommentchange}[2]{\ifnum\showcommentsflag > 0%
        \Red{\sout{#2}}\Red{#1}%
        \else
        #1
        \fi
}

\showcomments

\newlength{\abstwidth}
\begin{document}
% set sloppy attitude to line breaks
\sloppy
 
\pagestyle{empty}
 
\begin{flushright}
LU TP 22-02\\
MCnet-22-01\\
\end{flushright}

\vspace{\fill}

\begin{center}
{\Huge\bf Hyperfine splitting effects in string hadronization}\\[4mm]
{\Large Christian~Bierlich, Smita~Chakraborty, G\"osta~Gustafson, and Leif~L\"onnblad} \\[3mm]
{\texttt{christian.bierlich@thep.lu.se}, \texttt{smita.chakraborty@thep.lu.se}, \texttt{gosta.gustafson@thep.lu.se}, \texttt{leif.lonnblad@thep.lu.se}}\\[1mm]
{\it Theoretical Particle Physics,}\\[1mm]
{\it Department of Astronomy and Theoretical Physics,}\\[1mm]
{\it Lund University,}\\[1mm]
{\it S\"olvegatan 14A,}\\[1mm]
{\it SE-223 62 Lund, Sweden}
\end{center}

\vspace{\fill}

\begin{center}
\begin{minipage}{\abstwidth}
{\bf Abstract}\\[2mm]
	We revisit the recipe for hadron formation in the Lund string hadronization model. Given an incoming quark or quark-diquark pair, weights for hadron formation are updated to take hyperfine splitting effects arising from the mass difference between $\mrm{u,d}$-type and $\mrm{s}$-type quarks. We find that the procedure improves the description of hadron yields in \epem collisions and the $\phi$ cross section in neutral current DIS. We also show results for proton collisions, and discuss the future use of this study in the context of small system collectivity.
\end{minipage}
\end{center}

\vspace{\fill}

\phantom{dummy}

\clearpage

\pagestyle{plain}
\setcounter{page}{1}

\section{Introduction}
\label{sec:intro}

The surprising discovery from LHC \cite{ALICE:2016fzo} of continuous strangeness enhancement with final state multiplicity across collision systems,
has spurred a renewed interest in hadronization models. While increased strangeness production in heavy ion
collisions is traditionally seen as a signature of Quark--Gluon Plasma (QGP) formation \cite{Rafelski:1982pu,Koch:1982ij,Koch:1986ud}, microscopic hadronization models such as the Lund string model \cite{Andersson:1983ia} or the cluster model \cite{Webber:1983if} can describe similar effects using models of
cluster reconnections \cite{Gieseke:2017clv,Duncan:2018gfk} or string interactions, such as rope formation \cite{Bierlich:2014xba,Bierlich:2015rha} or junction formation \cite{Christiansen:2015yqa,Bierlich:2015rha}.
In the case of junction formation, even more attention has been gathered by the observation that also charm baryon yields
are enhanced in hadronic collisions compared to \epem \cite{ALICE:2020wfu}.

What all these approaches have in common, is their connection to the LEP baseline. As the basic notion of the models is that of
\textit{jet universality}, \ie that the same physics model should hold in all collision systems, parameters are often estimated
using data from \epem collisions at the $\mathrm{Z}$ pole \cite{Skands:2014pea}. This, in turn, means that no model attempting to describe
features of pp collisions, will perform better than the LEP baseline. And while \eg the Lund model as implemented in the \pythia
Monte Carlo event generator \cite{Sjostrand:2014zea} does a reasonable job at describing the data, there are still details less well described, even
total yields of some of the most intensely studied baryon and meson species in proton collisions, such as the $\phi$ meson and $\Omega^-$ baryon, both characterized by having the maximal amount of inner strangeness. 

Several tunes for the  \pythia event generator have been presented, as new experimental results have been available; the present default tune in \pythia 8 is the "Monash tune" from 2014 \cite{Skands:2014pea}. However, the basic parametrization is not changed at least since JETSET 6.2 in 1986 \cite{Sjostrand:1985ys}.

The production of different hadrons depends both on their quark content and the hadron mass. As discussed in \sectref{sec:fragmentation}, the string can break via $q\bar{q}$ pair production in a kind of tunnelling process \cite{Brezin:1970xf}. Here the virtual quark and antiquark are regarded as produced in a single point and pulled apart by the string tension until they come on shell, and the result is a Gaussian suppression for higher quark masses. 

The production probability also depends on the mass of the hadron, which is sensitive to the spin-spin interaction between two quarks, proportional to $\mathbf{M}_1\cdot\mathbf{M}_2$, where $M_i \propto g/\mu_i$ are the magnetic moments and $\mu_i$ the masses of the quark and the antiquark in a meson, or in any pair of two quarks in a baryon. This effect separates the mass of the $\rho$-meson and the pion, leading to a suppression of the $\rho/\pi$ ratio by about a factor 1/2 (on top of the factor 3 from spin counting). Due to the larger mass of the strange quark the mass difference between K$^*$ and K is smaller, and the K$^*$/K similarly less suppressed.
However, lacking experimental information at the time, the interaction between
two strange quarks in an ss diquark in a baryon, or in an $\mathrm{s\bar{s}}$ pair 
in a $\phi$ meson, did not get a corresponding extra reduced suppression, beyond the suppression from a single s-quark. 
(The $\eta$ and $\eta'$ mesons, both with a significant $\mathrm{s\bar{s}}$ content and 
with very low production rates, got their individual tunable suppression factors fitted to data;
see also the discussion in \citeref{Andersson:1994xd}.) In \sectref{sec:hyperfine}
below, we present a relatively simple way to take the modified spin interaction between
two strange quarks into account. It is most important for multi-strange baryons, but it has also a non-negligible effect on the production of $\phi$ mesons. The lower suppression of multi-strange baryons will also imply a larger suppression for  $\Delta^{++}$.

In the following we will first recap the basic features of string
hadronization in \sectref{sec:fragmentation}, before we introduce the improvements concerning the strange quarks in \sectref{sec:hyperfine}. A tune to the modified hadronization is presented in \sectref{sec:tuning}. We present some result in \sectref{sec:results}, noting that the majority of results presented here are either for \epem collisions or low-multiplicity proton collisions, in order to compare to an environment as free as possible from effects from rope hadronization, junction formation etc. Finally in \sectref{sec:conclusion} we present our conclusions.

\section{Lund string fragmentation}
\label{sec:fragmentation}

We will here discuss the basics of the Lund string hadronization model as implemented in \pythia; for more details see \citeref{Andersson:1983ia}.
In the model the confining colour field is approximated by a ``massless relativistic string''.
(For the dynamics of such a string we refer to \citeref{Artru:1979ye}.) A straight string is invariant under longitudinal boosts,
with a given energy per unit length (or tension) $\kappa \approx 1$ GeV/fm, but no longitudinal momentum. It can be visualised as a thin tube with a homogeneous electric field, similar to a vortex line in a superconductor (with electric and magnetic fields exchanged). 

The string can break by the production of a $\mrm{q}\bar{\mrm{q}}$ pair, in a process analogous to the production of \epem pairs in a homogeneous electric field. This gives \cite{Schwinger:1951nm}:
\begin{equation}
\exp(-\pi(\mu^2+p_\perp^2)/\kappaS) = \exp(-\pi \mu^2/\kappaS) \times \exp(-\pi p_\perp^2/\kappaS),
\label{eq:tunnel}
\end{equation}
where $\mu$ and $p_\perp$ are the mass and transverse momentum for the 
quark and anti-quark in the pair. The quark and anti-quark are then pulled in opposite directions
by the string tension. The result in \eqref{eq:tunnel}
can also be regarded as a result of tunnelling, where the quark and antiquark are produced 
in a single point as virtual particles, and pulled apart a distance $2 \sqrt{\mu^2 + p_\perp^2}/\kappaS$ until they can come on shell \cite{Brezin:1970xf}. A quark and an anti-quark from neighbouring breakups can combine to form a meson, if the mass is correct.

Another important feature of the model is that a gluon is treated as a momentum 
carrying ``kink'' on the string, pulled back by the force $2\kappaS$ from the two adjacent string pieces. The fragmentation of such a string with several intermediate gluons is discussed in \citeref{Sjostrand:1984ic}.
In this section we will, however, focus on the \textbf{fragmentation of a straight string} between a quark and an antiquark, without intermediate gluons.

\subsection{Meson production}
\label{sec:mesons}

\subsubsection{Ideal case with a single meson species}

\begin{figure}
\begin{center}
   \includegraphics[width=0.6\textwidth,bb=200 430 600 690,clip]{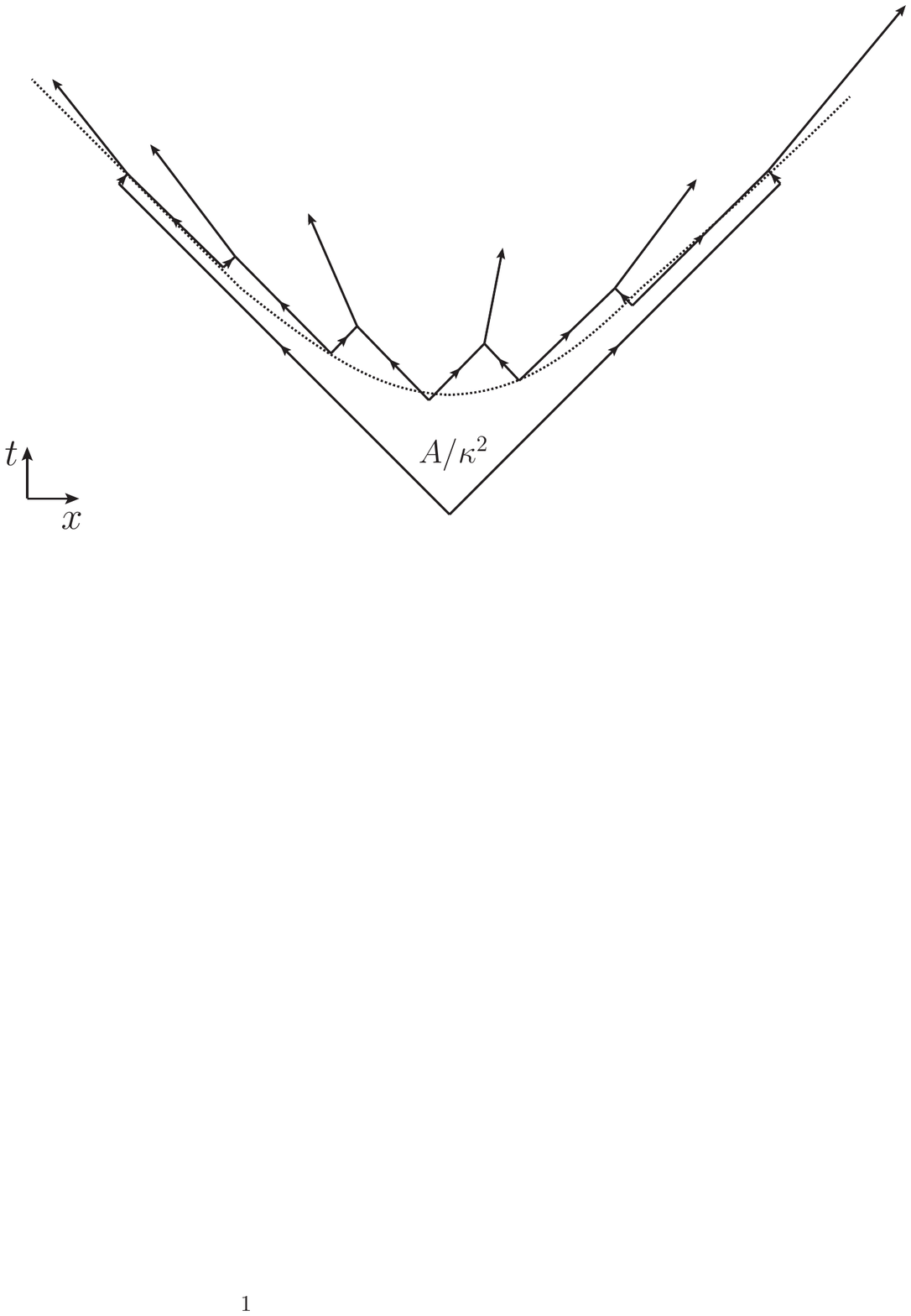}
\end{center}
  \caption{ \label{fig:fan}Breakup of a string between a quark and an
   anti-quark in a $x-t$ diagram. New $q\bar{q}$ pairs are produced
   around a hyperbola, and combine to the outgoing hadrons. The
   original $q$ and $\bar{q}$ move along light-like trajectories. The area
   enclosed by the quark lines is the coherence area $A$ in
  \eqref{eq:lu-had}, in units of the string tension $\kappaS$.}
\end{figure}

For simplicity we first limit ourself to the situation with a single hadron 
species, neglecting also transverse momenta. In this simplified situation the breakup to a state with $n$ hadrons is given by the expression \cite{Andersson:1983jt}: 
\begin{equation}
   \label{eq:lu-had}
    \diff{\mathcal{P}} \propto \prod_{i=1}^n \left[N\dtwo{p_i} \delta(p^2_i - m^2)\right]
    \delta^{(2)} \!\left(\sum p_i - P_{\mathrm{tot}}\right)\exp\left(-bA\right).
\end{equation}
Here $p_i$ (with $i = 1, ..., n$) and $P_{\mathrm{tot}}$ are two-dimensional vectors.
The expression is a product of a phase space factor, where the parameter $N$
expresses the ratio between the phase space for $n$ and $n-1$ particles, and
the exponent of the imaginary part of the string action,
$bA$. Here $b$ is a parameter and $A$ the space--time area covered by the
string before breakup (in units of the string tension $\kappaS$). 
This decay law can (for large enough energies) be implemented as an iterative 
process, where each successive hadron takes a fraction $z$ of the remaining 
light-cone momentum ($p^\pm = E \pm p_z$) along the positive or negative light-cone, depending on from which end the hadron is "peeled off". The values of these momentum fractions are then given by the distribution
\begin{equation}
    \label{eq:lu-frag}
    f(z) = N\frac{(1-z)^a}{z}\exp(-bm^2/z).
\end{equation}
Here $a$ is related to the parameters $N$ and $b$ in \eqref{eq:lu-had} by
normalization. (In practice $a$ and $b$ are determined from
  experiments, and $N$ is then determined by the normalization constraint.)
The result in \eqref{eq:lu-frag} is in principle valid for strings stretched between
partons produced in a single space--time point, and moving apart as illustrated 
in the space--time diagram in \figref{fig:fan}.
    
The expression in \eqref{eq:lu-had} is boost invariant, and the hadrons are
produced around a hyperbola in space--time. A Lorentz boost in the
$x$-direction will 
expand the figure in the $(t+x)$ direction and compress it in the $(t-x)$
direction  
(or \textit{vice versa}). Thus the breakups will be lying along the same
hyperbola, and  
low momentum particles in a specific frame will always be the first to be
produced in that special frame.  
The typical proper time for the breakup points is given by
\begin{equation}
    \label{eq:prod-time}
    \langle \tau^2 \rangle = \frac{1 + a}{b\kappaS^2}.
\end{equation}
With parameters $a$ and $b$
determined by tuning to data from \epem\ annihilation at LEP\footnote{The Monash tune \cite{Skands:2014pea} gives $a = 0.68$ and $b = 0.98\, \mathrm{GeV}^{-2}$.}, and $\kappaS$ equal to 0.9-1~GeV/fm, \eqref{eq:prod-time} gives a typical breakup time of 1.5~fm.

\subsubsection{Flavour, transverse momentum,  and spin}
\label{sec:flavourspin}

In practice it is also necessary to account for different quark and hadron
species, and for quark transverse momenta. 
With $\kappaS \approx$ 1 GeV/fm \eqref{eq:tunnel} implies that strange quarks are suppressed by roughly a factor 0.3 relative to a u- or a d-quark. It also means that the quarks are produced with an average $p_\perp \sim 250$ MeV, independent of its flavour.
When a u- and a $\bar{\mathrm{d}}$-quark from neighbouring breakups combine to form a meson, it can result in either a pion or a $\rho$-meson\footnote{In default \pythia only ground state hadrons, where the constituent quarks have relative angular momentum $L=0$, are included. This means the octet and singlet pseudoscalars and vector mesons, and the octet and decuplet baryons. 
The semiclassical picture, with a string piece and a quark and an anti-quark at the ends, which have average transverse momenta $\approx 250$ MeV, will give $\langle L \rangle \sim 0.1$, and therefore a contribution from about 10\% $L=1$ states may be expected.
An option which includes $L=1$ states is also available in the \pythia event generator. However, when 
these resonances have decayed, the result is very similar to the default version.}.
As stated in the introduction, the relative probabilities depend on the number of available spin states, but also by the normalization of the meson wavefunction affected by the spin-spin interaction between the quarks. The latter is proportional to $\mathbf{M}_1 \cdot \mathbf{M}_2$, where $M_i \propto g/\mu_i$ are the colour magnetic moments of the two components.
As discussed in \citeref{Andersson:1983ia}, this can compensate the factor 3 from spin counting, giving a relative probability closer to 1, in agreement with observations \cite{Cohen:1980zg, Skands:2014pea}. 

Of interest here is that the colour magnetic moment is proportional to $1/\mu$, where $\mu$ is the “effective” mass of the quark. Therefore this effect is smaller for hadrons with a strange quark, giving a $\mathrm{K}^*/\mathrm{K}$ ratio which is larger than the $\rho/\pi$ ratio. This effect was included already in early versions of \pythia. 
In conclusion the overall $q\bar{q}$ production rate is determined by the parameters $a$ and $b$ in \eqref{eq:lu-frag}, and the relative rates 
for $\pi$, $\rho$, K, and K$^*$, which are determined by three tuneable parameters, and which we for the purpose of this paper name as follows:

\begin{itemize}
	\item[$\rho$:] The overall suppression of $\mathrm{s\bar{s}}$ string breaks relative to $\mathrm{u}$ or $\mathrm{d}$ ones. (In \pythia settings this parameter is called \texttt{StringFlav:probStoUD}, and has a default value of 0.217).
	\item[$y_{\mathrm{ud}}$:] The relative production ratio of vector mesons to pseudoscalar mesons for $\mathrm{u}$ or $\mathrm{d}$ types, not accounting for the factor 3 from spin counting. (In \pythia settings this parameter is called \texttt{StringFlav:mesonUDvector}, and has a default value of 0.50).
	\item[$y_{\mathrm{s}}$:] The relative production ratio of vector mesons to pseudoscalar mesons for $\mathrm{s}$ types. (In \pythia settings this parameter is called \texttt{StringFlav:mesonSvector}, and has a default value of 0.55).
\end{itemize}
This implies that for \textit{e.g.} a $\mathrm{u}\bar{\mrm{d}}$ pair  becomes a $\pi^+$ or a
 $\rho^+$ with the following probabilities (the factor 3 comes from spin counting):
 \begin{equation}
 P_{\pi}= 1/(1+3y_{\mathrm{ud}}); \,\,\,\,\,\,\, P_{\rho} = 3y_{\mathrm{ud}}/(1+3y_{\mathrm{ud}}),
 \label{eq:mesonprob}
 \end{equation}
 and similar probabilities for K and $\mathrm{K}^*$, with $y_{\mathrm{ud}}$ exchanged to $y_{\mathrm{s}}$.
%with probability $1/(1+3y_{\mathrm{ud}})$ and a $\rho^+$ with probability $3y_{\mathrm{ud}}/(1+3y_{\mathrm{ud}})$, and similar probabilities for K and $\mathrm{K}^*$, with $y_{\mathrm{ud}}$ exchanged to $y_{\mathrm{s}}$.

The production of mesons with an $\mrm{s}\bar{\mrm{s}}$ pair is more complicated. 
For the isospin 0 vector mesons, $\phi$ is essentially a pure $\mrm{s}\bar{\mrm{s}}$ state, while
$\omega$ contains only non-strange quarks. Thus we have $\omega = (\mrm{u}\bar{\mrm{u}}-\mrm{d}\bar{\mrm{d}})/\sqrt{2}$ (with the iso-triplet
$\rho^0 =(\mrm{u}\bar{\mrm{u}}+\mrm{d}\bar{\mrm{d}})/\sqrt{2}$). 
The $\omega$ mass is approximately the same as for $\rho$, 
and their production rates are also approximately the same.
In default \pythia an $\mrm{s}\bar{\mrm{s}}$ pair will give a $\phi$ meson, with the same probability ($3y_{\mathrm{s}}/(1+3y_{\mathrm{s}})$) as a $\mathrm{K}^{*-}$ from an
$\mrm{s}\bar{\mrm{u}}$ pair,  although the $\phi$ has two strange quarks and K$^*$ only one.
In \sectref{sec:hyperfine-meson} we discuss how we can take this difference into account.

The isospin 0 pseudoscalars, $\eta$ and $\eta'$, are mixtures of the flavour singlet state 
$\eta_1 = (\mrm{u}\bar{\mrm{u}}+\mrm{d}\bar{\mrm{d}} +\mrm{s}\bar{\mrm{s}})/\sqrt{3}$
and the octet state 
$\eta_8 = (\mrm{u}\bar{\mrm{u}}+\mrm{d}\bar{\mrm{d}} -2\mrm{s}\bar{\mrm{s}})/\sqrt{6}$,    
with a mixing angle $\theta_P$ between -10 and -20 degrees:
\begin{eqnarray}
\eta &=& \cos(\theta_P)\, \eta_8 -\sin(\theta_P)\, \eta_1, \nonumber \\
\eta' &=& \sin(\theta_P)\, \eta_8 + \cos(\theta_P)\, \eta_1.
\label{eq:mixing}
\end{eqnarray}
In a chiral symmetry limit, with all quarks being massless, $\eta_8$ would be a pseudo-Goldstone particle, along with the pion and the kaon. However, as both $\eta$ and $\eta'$ are quite heavy, we are far from this limit. These particles also have quite low production rates. The solution in \pythia is to suppress them both with an individual tunable parameter. We note that this procedure makes the interpretation of the parameter $y_\mrm{s}$ as the vector to pseudoscalar ratio ambiguous, as if no meson is chosen, a new trial break-up is made.

\subsection{Baryon production}
\label{sec:baryons}

We here describe baryon production in a single string as realized in the 
``popcorn'' model, which is the default treatment of baryon production in \pythia.

In analogy with the production of mesons in \sectref{sec:mesons}, a
baryon--antibaryon pair can be formed if the string breaks by the
production of a diquark--antidiquark pair, forming a colour antitriplet and a
triplet respectively \cite{Andersson:1981ce}. 
In this case the $\mrm{B}\bar{\mrm{B}}$ pair will always have two quark flavours
in common and lie close to each other in momentum space.  Experimental
data from \epem annihilation show that this is not always
the case, and a model, with a stepwise production mechanism,
called the popcorn model, was presented in \citeref{Andersson:1984af}\footnote{A stepwise production mechanism was also suggested by Casher \emph{et al.} in ref.~\cite{Casher:1978wy}.}.
  
\begin{figure}
	\begin{center}
                \includegraphics[width=0.8\textwidth]{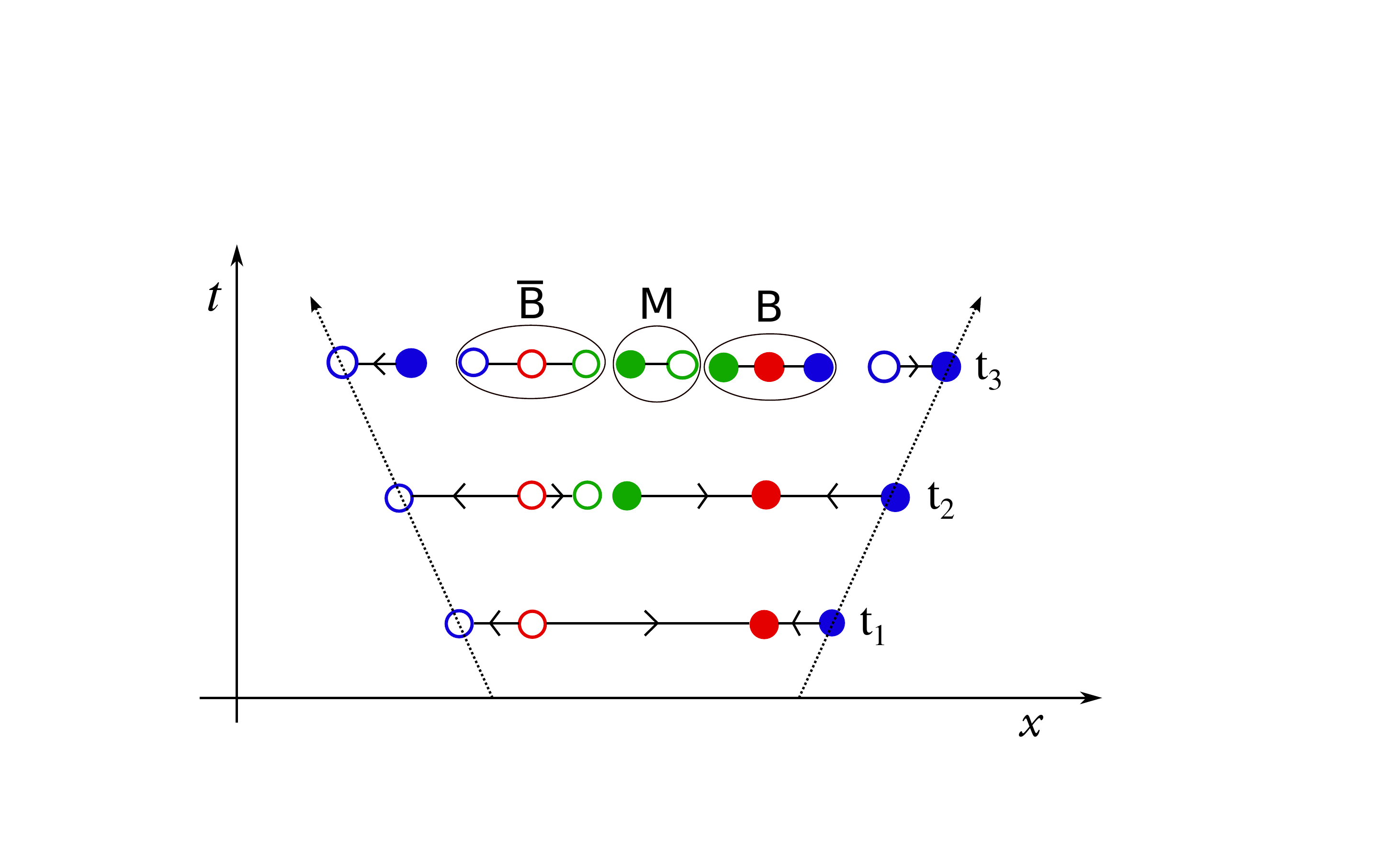}
                \caption{Illustration of the popcorn model with a string spanned 
                between a blue and an antiblue charge. At $t_1$ a red-antired
                $q\bar{q}$ pair is produced as a virtual fluctuation. In the green
                field between then a two green-antigreen pairs are produced at 
                $t_2$ and $t_3$. At $t_3$ also two new blue-antiblue pairs have 
                been produced closer to the original blue and antiblue charges. 
                The result is a baryon-antibaryon pair with a meson in between.}
                \label{fig:popcorn}
	\end{center}
\end{figure}

In a red--antired ($r\bar{r}$)
string-field, a $b\bar{b}$ pair can be produced as a vacuum
fluctuation. If the $r$ and $b$ charges form an antiblue antitriplet,
then new $\mrm{q}\bar{\mrm{q}}$ pairs can be produced in the green string field.
With a single such pair, the result would be similar to the
diquark--antidiquark production described above. However, if more than
one such pair is produced, we get not only a $\mrm{B}\bar{\mrm{B}}$ pair, but also
one or more mesons in between them, as seen \figref{fig:popcorn}. 
In this case the baryon and the
antibaryon may have only a single flavour in common. 
However, as the first $\mrm{q}\bar{\mrm{q}}$ pair (the blue pair above) is produced as a virtual 
fluctuation with limited lifetime, the production of several intermediate
mesons, or the more massive vector mesons, is strongly suppressed.
In the \pythia implementation therefore only a single intermediate
meson is allowed, with the overestimated $\rho$ and $\omega$ mesons 
simulating the neglected combinations of two or three pions.

The default \pythia~implementation of diquark production, has three relevant parameters in addition to the suppression for every strange quark compared to non-strange quarks:
\begin{itemize}
	\item[$\xi$:] suppression for a diquark compared to a single quark, independent of flavour 
	and spin. (In \pythia settings this parameter
	is called \texttt{StringFlav:probQQtoQ}, and has a default value of 0.081).
	\item[$x$]: extra suppression for diquarks with (any) strange content relative to diquarks without. (In 
	\pythia settings this parameter is called \texttt{StringFlav:probSQtoQQ}, and has a default
	value of 0.915.) This suppression is in addition to the normal suppression due to the presence of an s-quark ($\rho$), 
	\item[$y_\mrm{diq}$:] suppression of spin 1 diquarks relative to spin 0 diquarks. (In \pythia settings this parameter is
	called \texttt{StringFlav:probQQ1toQQ0}, and has a default value of 0.0275.) This suppression comes on
	top of the factor 3 enhancement of vector diquarks from counting spin states.
\end{itemize}
The production rate of each (single) diquark can thus be calculated as the product of a spin factor, a vertex factor and 
a single-diquark tunnelling factor. These can in turn be combined for each possible diquark pair. 

An essential feature here is that a baryon is an antisymmetric colour singlet state, and thus symmetric in flavour and spin. Therefore a diquark formed by two equal quarks can only have spin 1, and a ud-diquark must either have both spin and isospin 1, or both spin and isospin 0. 
The quark and diquark is then combined into a baryon, then again taking into account the total symmetry of the three-quark state. All relevant flavour information is determined by the parameters above, and in addition a baryon weight depending on the relevant \textbf{SU(6)} Clebsch-Gordan coefficients. As for the mesons, these may not necessarily add up to 1, and if no baryon is chosen a new trial is made.

\section{Accounting for hyperfine splitting effects}
\label{sec:hyperfine}

As discussed in \sectref{sec:fragmentation} the production rate for different 
hadrons is affected by the spin interaction of the quarks.
The spin-spin interaction between a quark and an anti-quark or between two quarks is proportional to the product of the two colour-magnetic moments, $\mathbf{M}_1 \cdot \mathbf{M}_2$, which gives the ratio 
3:(-1) between pairs with spin 1 and spin 0. As example this
reduces the $\rho/\pi$ ratio from 3 (as given by spin counting) by approximately 
a factor 0.5.
This effect is naturally related to the increased hadron mass due to the spin-spin 
interaction \cite{Andersson:1983ia}.
As this effect is due to the colour-magnetic moments of the quarks, the difference 
between ud-, us-, and ss-diquarks is given by relative factors 
$1/\mu_u^2 :  1/(\mu_u \mu_s): 1/\mu_s^2$, where 
$\mu_u$ and $\mu_s$ are unknown “effective” (constituent or current) quark 
masses. The same relative factors are obtained for the quark-antiquark pairs
u$\bar{\mrm{d}}$, u$\bar{\mrm{s}}$, and s$\bar{\mrm{s}}$.

As also discussed in \sectref{sec:fragmentation}, due to lack of experimental data at the time, 
in the implementation in \pythia
there is a reduced suppression for all quark pairs with at least one strange quark, but no 
further reduction for pairs with two strange quarks, neither for mesons nor for baryons.
For meson production a correction will enter into the assignment of meson species, when the ingoing
$q\bar{q}$ pair is given, \ie as corrections to the parameter $y_{\mrm{ud}}$ and $y_\mrm{s}$,
as introduced above. For baryon production, the effect enters into the relative weights given to
spin 1 diquarks ($y_\mrm{diq}$) and diquarks with strange content $x$. To keep the amount of parameters
constant, we will in the following set $x=1$ (thus removing the possibility of further suppression),
and focus only on $y_\mrm{diq}$.

\subsection{Modified meson production}
\label{sec:hyperfine-meson}

We choose a simple ansatz for including the hyperfine splitting (HFS) effect. As explained in \sectref{sec:flavourspin},
in default \pythia 
%parameters are already in place governing 
the relative rates of vector to pseudoscalar mesons is determined by the two parameters  $y_\mrm{ud}$ and $y_\mrm{s}$, where the latter determines the suppression of both 
$\mrm{K}^*$ and $\phi$. We now want to introduce a more flexible formalism, which allows a suppression which depends on the number of strange quarks.
To keep the number of parameters unchanged, we re-parametrize the suppression of vector mesons  as a common parameter $y_\mrm{m}$, which depends on the number of strange quarks, $n_\mrm{s}$, in the meson, 
%which determines the overall vector to pseudoscalar meson ratio, as a function of 
as follows:

\begin{equation}
	y_\mrm{m}( n_\mrm{s})% \equiv \frac{\mathcal{P}_\mrm{v}}{\mathcal{P}_\mrm{ps}} 
	= y_{\mrm{m}1} + n_\mrm{s} y_{\mrm{m}2}.
\end{equation}
Here $y_{\mrm{m}1}$ and  $y_{\mrm{m}2}$ are two new parameters. In order to obtain the same rate for the vector 
mesons $\rho$ and $\omega$, with only non-strange quarks, 
one can consider $y_{\mrm{m}1}  = y_\mrm{ud}$.

The important difference between this ansatz and the previous parametrization, is 
 instead for $\mrm{K}^*$ and $\phi$ mesons. While they were equally suppressed before,
 $\mrm{K}^*$ will be suppressed by a factor $y_\mrm{m1} + y_\mrm{m2}$, while $\phi$ now is less suppressed by the larger factor $y_\mrm{m1} + 2y_\mrm{m2}$. Remember that, as discussed in \sectref{sec:flavourspin},
$\eta$ and $\eta'$ are further suppressed by two individual parameters. Therefore the factor 
$y_\mrm{m}$ does not give the ratio between the vector meson $\phi$ and a corresponding isoscalar pseudoscalar meson. 
It rather gives the probability $3y_{\mathrm{m}}/(1+3y_{\mathrm{m}})$ for an s$\bar{\mrm{s}}$ 
pair to form a vector meson $\phi$. Thus the production of the pseudoscalars $\eta$ and $\eta'$ will not be affected; their individual suppression factors will just be rescaled to give the same production probabilities as before. As the probabilities do not add up to 1, if no meson is chosen a new trial is made.

\subsection{Modified baryon production}
\label{sec:hyperfine-baryon}

As explained in \sectref{sec:baryons}, baryon production is governed mainly by the production of diquarks in
the popcorn model, where two parameters $x$ and $y_\mrm{diq}$ governs the diquark rates.
The production rate of each (single) diquark is calculated as the product of a spin factor, a vertex factor and 
a single-diquark tunnelling factor where the two parameters enter. These can in turn be combined for each possible diquark pair.
We set $x=1$ to keep the number of tunable parameters unchanged, and modify $y_\mrm{diq}$ according to the same simple ansatz
as above for mesons. Thus $y_\mrm{diq}$ is redefined as:
\begin{equation}
	y_\mrm{diq} \equiv \frac{\mathcal{P}_{\mrm{v}}}{\mathcal{P}_{\mrm{s}}} = y_\mrm{d1} + n_\mrm{s} y_\mrm{d2}.
\label{eq:magnmoments}
\end{equation}

Technically, this modification enters only in the single-diquark tunnelling factors, and it suffices therefore
to modify those. The default and updated tunnelling factors are then given by (written as 
(tunnelling factor(s)) $=$ (default expression) $\mapsto$ (updated expression)):
\begin{eqnarray}
	\mathcal{P}(\mrm{ud}_1) &= \mathcal{P}(\mrm{uu}_1) = \sqrt{y_\mrm{diq}} \mapsto \sqrt{y_\mrm{d1}}, \label{eq:diqud0}\\
	\mathcal{P}(\mrm{us}_0) &= \sqrt{x} \mapsto \sqrt{x},\\
	\mathcal{P}(\mrm{su}_0) &= \sqrt{\rho x} \mapsto \sqrt{\rho x},\\
	\mathcal{P}(\mrm{us}_1) &= \sqrt{y_\mrm{diq}}\mathcal{P}(\mrm{us}_0) \mapsto \sqrt{y_\mrm{d1} + y_\mrm{d2}}\mathcal{P}(\mrm{us}_0),\\
	\mathcal{P}(\mrm{su}_1) &= \sqrt{y_\mrm{diq}}\mathcal{P}(\mrm{su}_0) \mapsto \sqrt{y_\mrm{d1} + y_\mrm{d2}}\mathcal{P}(\mrm{su}_0),\\
	\mathcal{P}(\mrm{ss}_1) &= \sqrt{\rho y} x \mapsto \sqrt{\rho(y_\mrm{d1} + 2y_\mrm{d2})} x. \label{eq:diqss1}
\end{eqnarray}
Note that the two quarks in a diquark must be symmetric in spin and flavour, 
just like the three quarks in a baryon.
Therefore the uu-, dd-, and ss-diquarks always have spin 1, the $\mrm{ud}_1$ diquark
has both spin \emph{and} isospin 1, and $\mrm{ud}_0$ has spin \emph{and} isospin 0.
Note that the apparent difference between \eg $\mathcal{P}(\mrm{us}_1)$ and $\mathcal{P}(\mrm{su}_1)$ is due to which quark is considered to be produced ``first'' in the model. The missing factor $\sqrt{\rho}$ appears in the vertex factor, and the resulting production probability turns out equal. 

We also note that a baryon in the spin 1/2 octet can always be produced by adding 
a quark to a spin 0 diquark. Therefore this modification,
when compared to standard \pythia, is in particular important for the decuplet, 
where it enhances $\Omega^-$ production and suppresses $\Delta^{++}$.

In the our implementation we introduce the new parameter $c$, which defines the two new parameters from the old one:
\begin{equation}
	y_\mrm{d1} = (1 - c)y_\mrm{diq}\mrm{,~~~~and~~~~}y_\mrm{d2}=c y_\mrm{diq}.
\label{eq:spinshare}
\end{equation}
This choice means that $y_\mrm{diq}$ now denotes the suppression for a diquark with a single s-quark, while $c$ describes the enhanced suppression of nonstrange diquarks and the reduced suppression of doubly strange diquarks.

\section{Tuning and \epem results}
\label{sec:tuning}

In this section we will describe estimation of parameters of the updated fragmentation model. Such a ``tuning'' process is to assure that the altered
model has at least as good a global description of data as the previous model. A modification such as the one introduced here
would be unsuccessful if it turned out that it could only improve some aspects of descriptions of data, by making
others worse. To keep a somewhat limited scope, we do, however, not perform a full retuning of all fragmentation parameters, or parameters related to final state radiation. Instead we only re-tune parameters directly affected by the performed changes, and stick to defaults \cite{Skands:2014pea}, the so-called ``Monash 2013'' tune for the rest. The resulting set of parameters can thus not be taken as a new 
``full tune'', but serve as reassurance that this model addition improves the overall description, in addition to providing
a reasonable set of parameters.
The program Rivet \cite{Bierlich:2019rhm} is used for data comparison, and Apprentice \cite{Krishnamoorthy:2021nwv} for post-processing. We are, however, not relying on a global minimization from Apprentice, but rather use it as a guiding hand.

The main \epem data used for the model validation, is hadron multiplicities obtained at the $\mrm{Z}$ pole, as described in the introduction. Since reported values are not always consistent with each other and PDG \cite{ParticleDataGroup:2020ssz}, we will in selected cases we give the published numbers some further attention in the following:

\begin{itemize}
        \item The total charged multiplicity is taken as a combination of data\footnote{In cases where we have performed a data combination ourselves, the average of data from different sources is taken as the normal weighted mean $\mu = \frac{\sum x_i/\sigma_i^2}{\sum 1/\sigma_i^2}$ with the error on the weighted mean $\sigma^2 = \frac{1}{\sum 1/\sigma_i^2}$.} from ALEPH \cite{ALEPH:1996oqp}, MARKII \cite{Abrams:1989rz}, OPAL \cite{OPAL:1998arz} and DELPHI \cite{DELPHI:1998cgx}.
	\item Meson multiplicities $\mrm{K}^\pm$ and $\mrm{K}^0_\mathrm{s}$ are taken from PDG \cite{ParticleDataGroup:2020ssz}.
	\item The $\mrm{K}^{*0}$ and $\mrm{K}^{*\pm}$ multiplicities from PDG \cite{ParticleDataGroup:2020ssz} are compatible with the individual measurements from DELPHI \cite{DELPHI:1994qgk,DELPHI:1996xro}, ALEPH \cite{ALEPH:1996oqp,ALEPH:1995njw} and OPAL \cite{OPAL:1992anp,OPAL:1997vmw}. We re-use the PDG average.
	\item The measured $\phi$ multiplicities at LEP and SLD are, as also noted in ref. \cite{Skands:2014pea} mutually discrepant. We take a very conservative approach, and use the envelope of the reported results by OPAL \cite{OPAL:1998vgx}, ALEPH \cite{ALEPH:1996oqp}, DELPHI \cite{DELPHI:1996xro} and SLD \cite{SLD:1998coh}, meaning that $\phi$ is not really given much constraining power.
	\item As mentioned above, the $\eta$ and $\eta'$ have special parameters to ensure their individual suppression. They are therefore not given any particular attention, but PDG values are included (without constraining power) for completeness.
	\item The proton multiplicity is given particular attention. The PDG average of $1.050 \pm 0.032$ per $Z$-event seems dominated by the result by SLD \cite{SLD:2003ogn} ($1.054 \pm 0.035$). The SLD result is slightly higher than results by ALEPH \cite{ALEPH:1997jlh} ($1.00 \pm 0.07$), OPAL \cite{OPAL:1994zan} ($0.92 \pm 0.11$), but agrees with DELPHI \cite{DELPHI:1995kfu}, which however does have a large error ($1.07 \pm 0.14$). If we exclude SLD from the average, we obtain a proton multiplicity per $Z$-event of $0.99 \pm 0.05$, which is the value we use. We cross check against the SLD value not extrapolated to full phase space.
        \item As noted in ref. \cite{Skands:2014pea}, the published measurements of $\Delta^{++}$ are mutually discrepant by $2\sigma$. We re-use the conservative average with increased errors from that reference ($0.09 \pm 0.017$).
        \item The $\Lambda^0$, $\Xi^\pm$ and $\Omega^-$ multiplicities are taken from OPAL \cite{OPAL:1996gsw}, which also correspond well with (and in fact drives) the PDG averages.
\end{itemize}

\begin{figure}
	\begin{center}
                \includegraphics[width=0.45\textwidth]{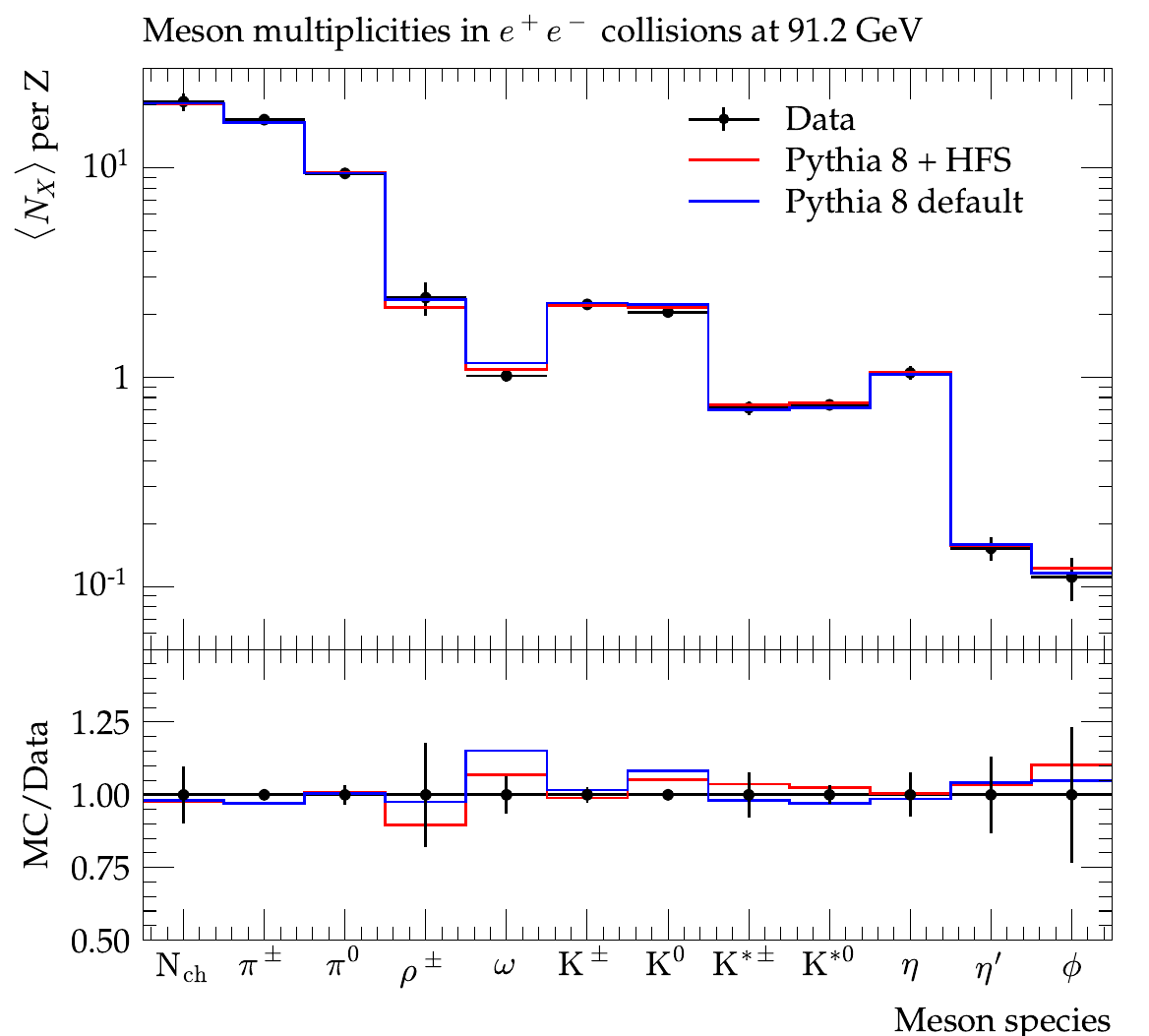}
                \includegraphics[width=0.45\textwidth]{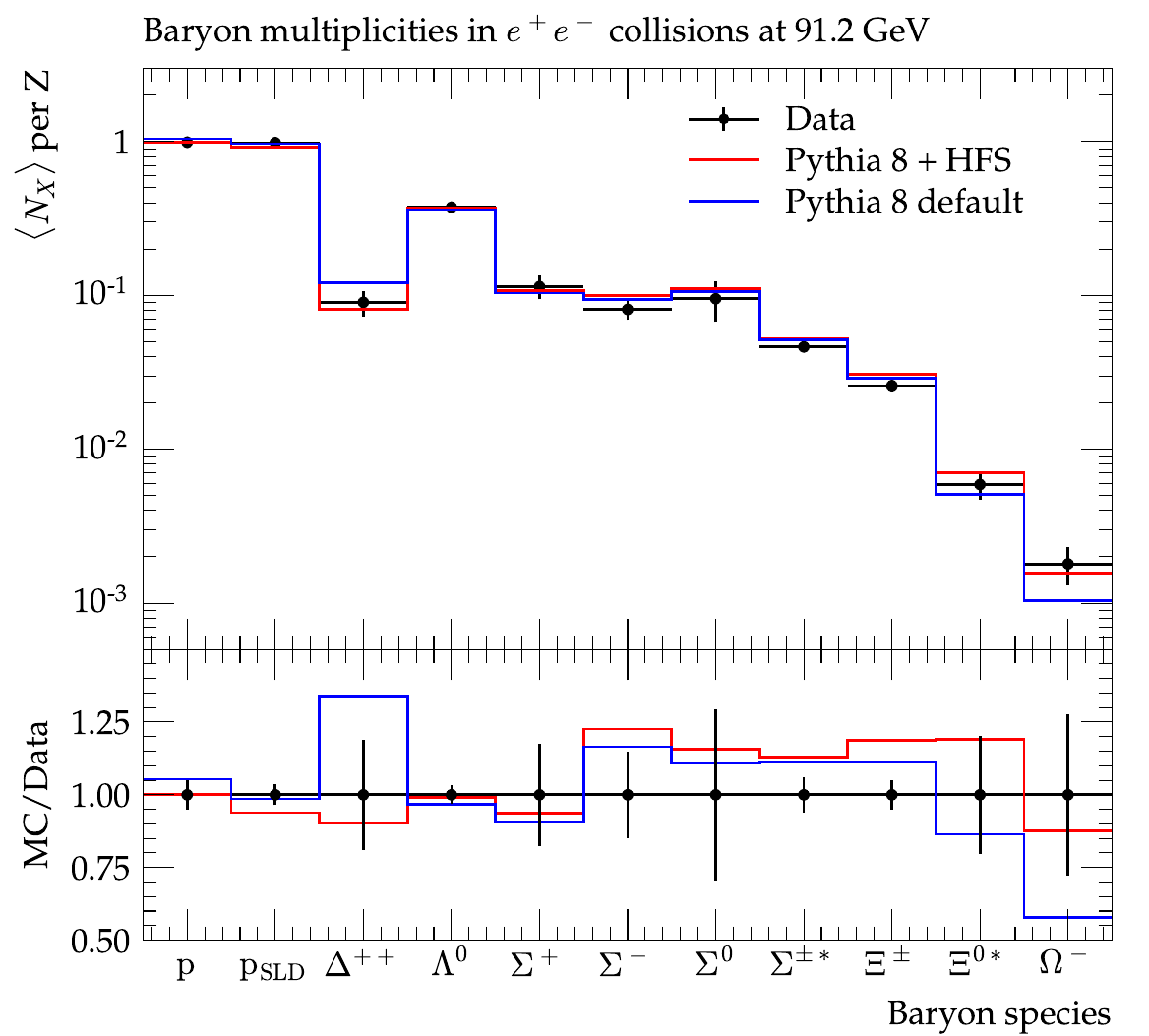}
		\caption{Yields of identified particles (per $\mrm{Z}$) in \epem
		 collisions at the $\mrm{Z}$ pole.
		Mesons to the left, and baryons to the right. The \pythia default tune (in blue) is compared with the model extension with reasonable parameters (in red), as described in the text. Data are from various LEP experiments and SLD, with references in the text.}
                \label{fig:ee-yields}
	\end{center}
\end{figure}

In \figref{fig:ee-yields} we show hadron yields per $\mrm{Z}$ boson from \epem collisions at the $\mrm{Z}$ pole. The model addition based on HFS is shown in red, and compared to default \pythia in blue.
 There are many parameter combinations which can describe \epem yields at a reasonable level, not least because the experimental error bars on the rare mesons and baryons, which are most sensitive to the presented model additions, are rather large. Hence, it also makes little sense to perform the tuning as a global $\chi^2$ minimization or similar. The strategy used is as follows, resulting parameters are listed in table \ref{tab:parameters}:
\begin{itemize}
\item Event shape observables will not change, as the parton shower is left untouched, and Lund fragmentation function parameters $a$ and $b$ are fixed to default values. Also the total charged or $\pi^\pm$ multiplicity is left almost unchanged.
	\item A value of the parameter $\rho$ is chosen, such that kaons from the pseudoscalar nonet is reproduced at least as well as in the default tune.
	\item A value of the parameter $\xi$ is chosen, such that our updated proton multiplicity is well reproduced. In \figref{fig:ee-yields} (right) we show the proton multiplicity from SLD next to it for comparison.
	\item Values for parameters $y_\mrm{m1}$ and $y_\mrm{m2}$ are chosen such that all recorded mesons in the vector nonet are now reproduced within error bars.
	\item Parameters for $\eta$ and $\eta'$ suppression are chosen to reproduce the yields form the default tune.
	\item Parameters $y_\mrm{diq}$ and $c$ are chosen to give a reasonable description of $\Delta^{++}$ and $\Omega^-$. 
\end{itemize}

\begin{table}
\begin{center}
\begin{tabular}{lccr}
        \hline
        \cline{1-4}
	Parameter name & \pythia name  & Default value & Retuned value \\
        \hline
	$\rho$ & \texttt{StringFlav:probStoUD} & 0.21 & 0.19 \\
	$\xi$ & \texttt{StringFlav:probQQtoQ} & 0.81 & 0.072 \\
	$y_\mathrm{diq}$ & \texttt{StringFlav:probQQ1toQQ0} & 0.0275 & 0.04 \\
	$x$ & \texttt{StringFlav:probSQtoQQ} & 0.915 & 1 (fixed) \\
	- & \texttt{StringFlav:etaSup} & 0.60 & 0.55 \\
	- & \texttt{StringFlav:etaPrimeSup} & 0.12 & 0.11 \\
	- & \texttt{StringFlav:mesonUDvector} & 0.50 & - \\
	- & \texttt{StringFlav:mesonSvector} & 0.55 & - \\
	$c$ & - & - & 0.65 \\
	$y_\mathrm{m1}$ & - & - & 0.4 \\
	$y_\mathrm{m2} $ & - & - & 0.25 \\
        \hline
        \cline{1-4}
\end{tabular}
\end{center}
	\caption{\label{tab:parameters}Table of parameters for the default (Monash 2013 \cite{Skands:2014pea}) and updated fragmentation model, listed with parameter names as given in this manuscript in the first column, \pythia names (when applicable) in the second column, followed by the default and retuned values. As it can be seen directly from the table, the number of free parameters in the default and the updated models are the same.}
\end{table}

\section{Results for ep and pp}
\label{sec:results}
In this section we go on to study \ep~and pp collisions, using the updated fragmentation model with parameters listed in table \ref{tab:parameters}. We first note that while some effects are expected, the default \pythia description already does a reasonable job for most relevant observables, so enormous effects are not expected. This is also not the point. As indicated in the introduction, the main goal of this paper is to establish a more thorough baseline for further studies.

\begin{figure}
	\begin{center}
                \includegraphics[width=0.45\textwidth]{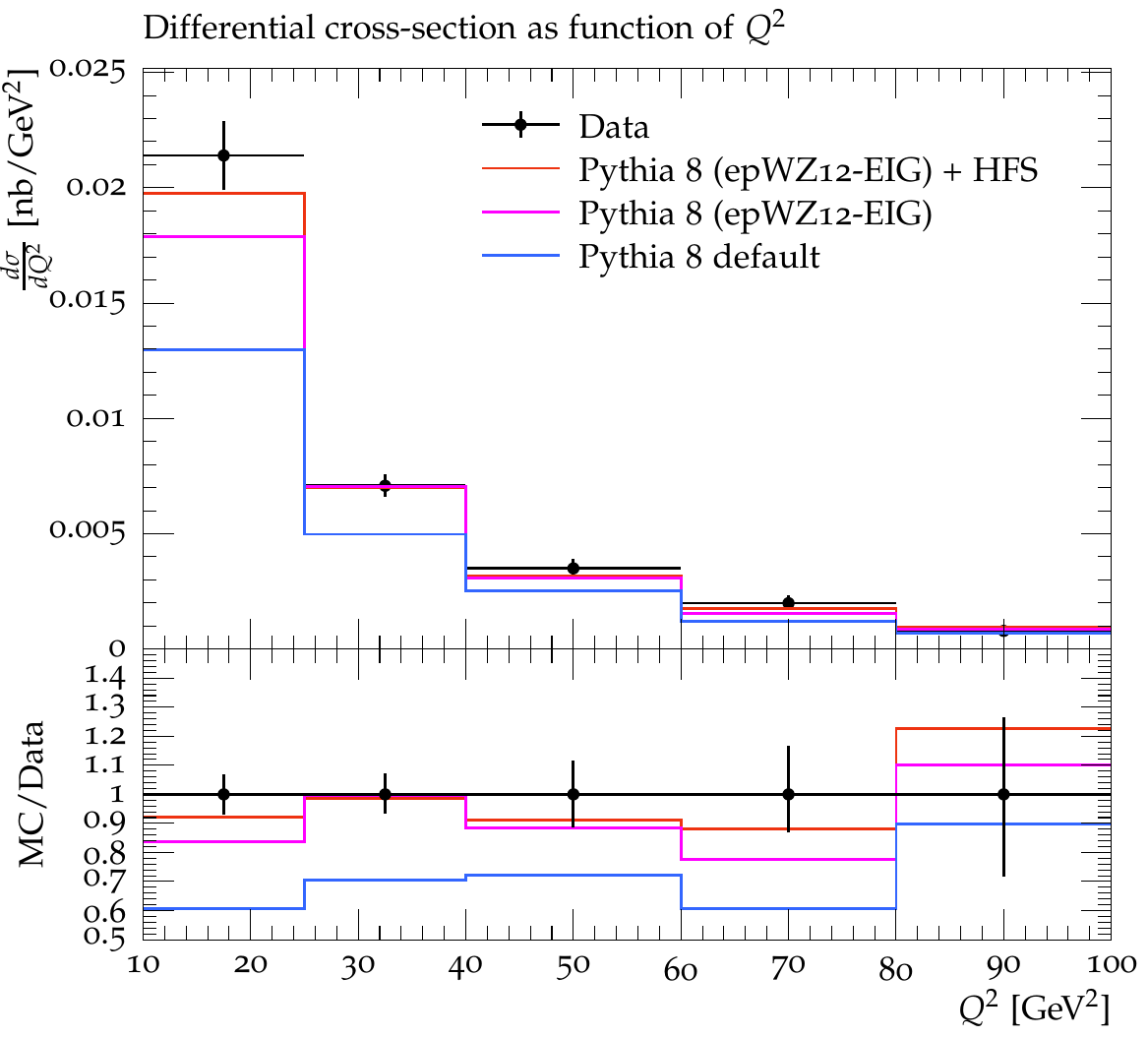}
                \includegraphics[width=0.45\textwidth]{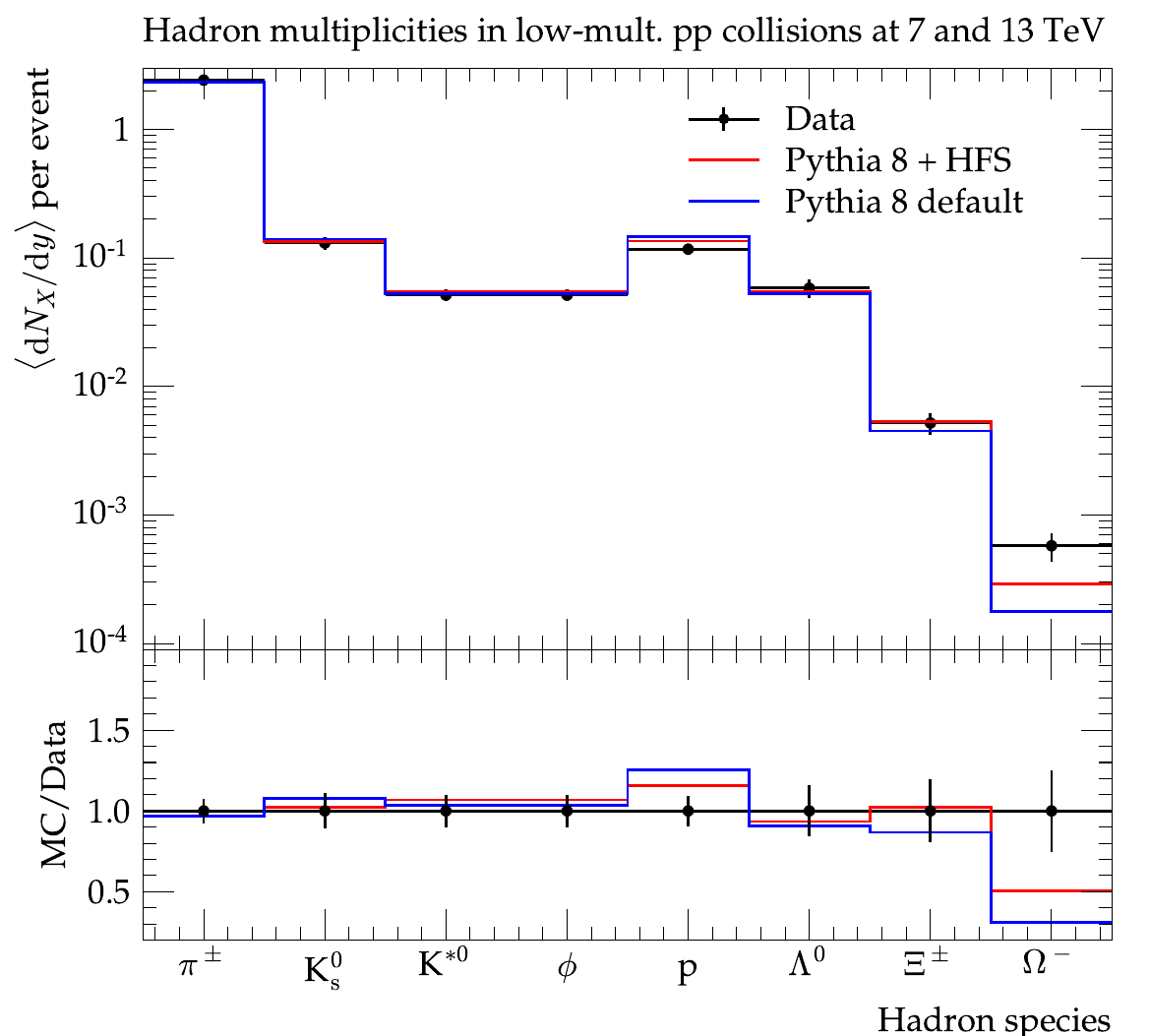}
		\caption{To the left we show $\phi$ production as function of $Q^2$ in neutral current DIS, as measured by ZEUS \cite{ZEUS:2002rhr}, compared to \pythia default (blue), \pythia with the epWZ12-EIG PDF (magenta) and finally with the hyperfine splitting effect (HFS) added on top (red). To the right we show hadron yields measured by ALICE in low multiplicity pp collisions at 7 TeV ($\pi^\pm$, $\mrm{K^0_s}$, $\mrm{p}$, $\Lambda^0$, $\Xi^\pm$ and $\Omega^-$) \cite{ALICE:2016fzo} and 13 TeV ($\mrm{K}^{*0}$ and $\phi$) \cite{ALICE:2019etb}. The data is again compared to \pythia default (blue) and with the HFS effects added (red).}
                \label{fig:ep-pp-results}
	\end{center}
\end{figure}

In \ep~collisions, results from ZEUS \cite{ZEUS:2002rhr} from 2002 on inclusive $\phi$ production in neutral current DIS, showed a large enhancement with respect to expectation. The most stunning result from that paper is arguably the $\phi$ cross section in the target region at small Bjorken $x$ (less than $\approx 0.006$), which is more than a factor of 2 above expectation. Indeed this result was taken as a clear indication of an enhanced strange sea in the proton at small $x$. As the \pythia DIS description is generally not capable of correctly describing basic particle multiplicities at small $x$ in the target region, we have to compare to some of the less spectacular results of the paper, still showing the same trend. In \figref{fig:ep-pp-results} (left) we show a comparison to the integrated (over $2\cdot 10^{-4} < x < 10^{-2}$) $\phi$ cross section as function of $Q^2$. In blue we show \pythia default, which performs similarly as the HERWIG 5.1 Monte Carlo event generator \cite{Marchesini:1991ch} did in the original paper. Somewhat better agreement was achieved by the LEPTO 6.5 \cite{Ingelman:1996mq} and ARIADNE 4 \cite{Lonnblad:1992tz} models, but only after an artificial decrease of the model parameters corresponding to the \pythia parameter $\rho$ (see \sectref{sec:flavourspin}). Accounting for hyperfine splitting alone cannot bring \pythia in agreement with this data. In magenta we show the effect of changing the parton distribution function (PDF, through LHAPDF6 \cite{Buckley:2014ana}) to one obtained by the ATLAS collaboration \cite{ATLAS:2012sjl}, where sensitivity to the strange quark density of protons on $\mathrm{W}^\pm$ and $\mathrm{Z}$ production at small $x$ is exploited. On top of this effect, we show (in red) the effect of hyperfine splitting, which makes the calculation compatible with data.

Going to pp collisions in \figref{fig:ep-pp-results} (right), we show comparisons to ALICE data at low forward multiplicity\footnote{The data is, with inspiration from heavy ion collisions, divided into multiplicity classes based on forward production. We compare to the lowest multiplicity class only.}, at $\sqrt{s} = 7$ TeV \cite{ALICE:2016fzo} and 13 TeV \cite{ALICE:2019etb} respectively. At higher multiplicity we would expect rope effects (see \sectref{sec:intro}) to dominate, in particular for multi-strange baryon production, which is why we limit ourselves to the lowest multiplicity class. Even the lowest multiplicity class is, however, not completely free from overlapping strings, but as close as one can get in a \pp collision. We note that there is about a factor of six between the \epem results in \figref{fig:ee-yields} and the pp results in \figref{fig:ep-pp-results} (right). Accounting for the different rapidity ranges, a factor of two remains. In the string model, this gives the direct interpretation that these events corresponds to two strings, meaning exchange of a single gluon or Pomeron. The exception to this argument is $\Omega^-$ production, on which we comment separately below.

All hadron multiplicities are described to the same level or better than \pythia default. There are two interesting points to be mentioned, pointing towards future work. 
First of all, we see that even with retuning to the lower proton yield in \epem, \pythia still overshoots the proton multiplicity in low multiplicity pp collisions, giving apparent tension between the description of pp and \epem. This leaves two possibilities. Either one of the data sets cannot be trusted -- we re-iterate that the \epem results also had notable internal tension -- or a key physics ingredient is missing. In the first case, future studies could consider simply leaving the proton yields out of the \epem tuning efforts, and instead tune to low multiplicity pp.
Secondly, the $\Omega^-$ yield is still not well described, even with the baseline changed. This marks an opportunity for future studies of the rope model. We re-iterarate the argument above, that the low-multiplicity collisions should correspond roughly to two strings, leaving room for string overlap. We remark that also previous literature \cite{ALICE:2016fzo} showed an effect of rope formation for $\Omega^-$ production, even in the lowest multiplicity bin\footnote{We note that the lowest multiplicity bin for $\Omega^-$ is wider than for other species, leaving the possibility that more strings can be present in this event class open. This does, however, not change the main conclusion.}. Our result confirms that even with further attention given to the \epem baseline, there is still room for collective effects in low multiplicity pp. This is potentially a game-changing realization, as it is common for experimental analyses of small system collectivity, to use low multiplicity collisions as a no-collectivity baseline. This result suggests that the common strategy might be flawed. We will follow this point up in a future paper.

\section{Conclusion}
\label{sec:conclusion}

The Lund string model and its basic recipe for string breakings and formation of hadrons from the ingoing quarks and anti-quarks, has been a cornerstone of the \pythia Monte Carlo event generator, and provided the basis for many phenomenological and experimental studies in \epem, \ep, and pp, mostly in an unchanged form since it's first versions in the 1980s. In this paper we have updated the model to account for the impact of light quark mass differences, giving rise to modifications in the hadronic wave functions. With a simple ansatz, and no further parameters added to the model, we manage to improve in particular the description of $\rho^\pm$, $\omega$, $\Delta^{++}$ and $\Omega^-$ yields in \epem collisions. With this update, as well as by using updated PDFs, it is also possible to give a reasonable description of integrated $\phi$ production in neutral current DIS, and the updated baseline makes future precision studies of strangeness enhancement in pp using the rope hadronization model possible.

\section*{Acknowledgements}
This work was funded in part by the Knut and Alice Wallenberg
foundation, contract number 2017.0036, Swedish Research Council, contracts
number 2016-03291, 2017-0034 and 2020-04869, in part by the European
Research Council (ERC) under the European Union’s Horizon 2020
research and innovation programme, grant agreement No 668679, and in
part by the MCnetITN3 H2020 Marie Curie Initial Training Network,
contract 722104.

\bibliographystyle{utphys}
\bibliography{bibliography,aux-bib}

\end{document}